\documentclass[10pt,twocolumn,english,conference]{IEEEtran}
\usepackage[T1]{fontenc}
\usepackage[latin9]{inputenc}
\usepackage[a4paper]{geometry}
\usepackage{color}
\usepackage{float}
\usepackage{units}
\usepackage{amsmath}
\usepackage{amssymb}
\usepackage{graphicx}

\makeatletter

\providecommand{\tabularnewline}{\\}
\usepackage{times}\usepackage{epsfig}\usepackage{amsfonts}

\makeatother

\usepackage{babel}

\makeatother

\usepackage{babel}

\makeatother

\usepackage{babel}

\usepackage{url}

\graphicspath{{./Immagini/}}

\usepackage{graphicx}
\usepackage{caption}
\usepackage{cite}

\usepackage{fancyhdr}
\makeatother

\usepackage{babel}
\begin{document}
\title{Sensing Performance of Multi-Channel RFID-based Finger Augmentation
Devices for Tactile Internet}
\author{\IEEEauthorblockN{F. Naccarata, G. M. Bianco and G. Marrocco}}
\maketitle
\begin{abstract}
Radiofrequency finger augmentation devices (R-FADs) are a recently
introduced class of epidermal radiofrequency identification (RFID)
sensor-tags attached to the fingers, communicating with a body-worn
reader. These devices are promising candidates to enable Tactile Internet
(TI) applications in the short term. R-FAD based on auto-tuning RFID
microchips can be used as dielectric probes for the material of touched
objects. However, due to the nearly unpredictable intrinsic variability
of finger-object interaction, a single sensorized finger (single-channel
device) is not enough to guarantee reliable data sampling. These limitations
can be overcome by exploiting a multi-channel R-FAD sensorizing multiple
fingers of the hand. In this paper, the dielectric-sensing performance
of a multi-channel R-FAD, composed of sensors encapsulated into soft
elastomers, is numerically and experimentally characterized, involving
a set of volunteers. The inter-sensor coupling is negligible, thus
enabling simultaneous independent dielectric measurements. \textcolor{black}{The
multi-sensor configuration allows for 100\% reliability of the on-hand
communication link} for touched objects in a wide range of permittivity.
Experiments moreover demonstrate that multi-channel measurements can
halve the measurement uncertainty of the single-channel case. The
achievable precision is suitable to discriminate among low-, medium-,
and high-permittivity materials.
\end{abstract}

\begin{IEEEkeywords}
Auto-tuning antennas, Body-area internet of things, Epidermal sensors,
Flexible electronic, Radiofrequency identification.
\end{IEEEkeywords}

\section{Introduction}

\captionsetup[figure]{labelfont={small},font={small},labelformat={default},labelsep=period,name={Fig.}} \captionsetup[table]{labelfont={default},labelformat={default},labelsep=newline,name={TABLE},textfont={sc},justification=centering}{\let\thefootnote\relax\footnotetext{Work funded by Italian Ministry of University and Research, project DUAL SKIN. Ref. FISR 2020 COVID: FISR2020IP\_00227.}}{\let\thefootnote\relax\footnotetext{This paper has supplementary downloadable material available at http://ieeexplore.ieee.org, provided by the authors. This material includes a movie clip that shows the stretchability and conformability of the fingertip sensors. This material is 31.6 MB in size.}}{\let\thefootnote\relax\footnotetext{The authors are with the Pervasive Electromagnetics Lab, University of Rome Tor Vergata, Rome, Italy. Corresponding author: G. M. Bianco (Giulio.Maria.Bianco@uniroma2.it)}}

Finger augmentation devices (FADs) are finger-worn electronic tools
conceived to provide supplemental capabilities through an additional
augmentation other than their form \cite{Shilkrot15}. FADs can have
different form factors, spanning from proximal-phalanx rings \cite{Wilhelm20}
to nail-pads \cite{Yin21,Lee20}, and several possible applications,
like driving vehicles \cite{Bilius20,Gheran20} or controlling aerial
drones \cite{Yau20}. The most ambitious goal of FADs is the achievement
of a sensorial\emph{ ultrability}, namely providing the user with
an artificial sensation to either restore a lost or degraded natural
ability or even to add a new sense, going beyond the natural physical
feelings of humans \cite{Shilkrot15bis}.

\begin{figure}[t]
\begin{centering}
\includegraphics[width=8.2cm]{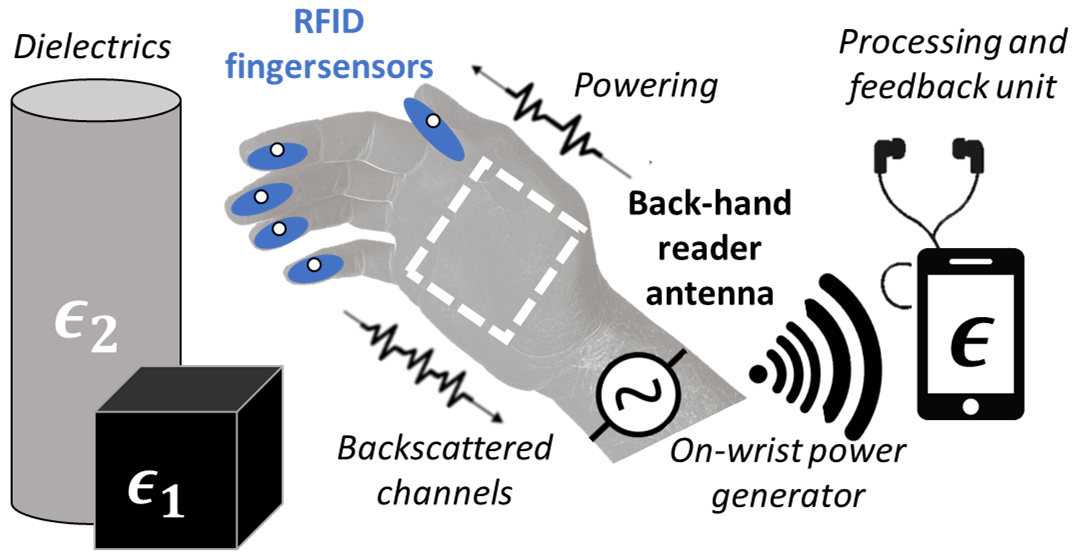}
\par\end{centering}
\caption{Pictorial concept of a dielectric sensing multi-channel R-FAD. \label{fig:Scheme-of-a}}
\end{figure}

FADs based on radiofrequency identification (RFID) were proposed to
help visually impaired people or to query tagged items \cite{Sedighi21,Singh21,Taylor19,Liang19}\textbf{.}
\textcolor{black}{Tighter integration with flexible sensors}\textcolor{blue}{{}
}directly attached over the fingers' skin denoted as R-FADs (radiofrequency
finger augmentation devices), have been introduced in \cite{Amendola19Numerical,Amendola19Application,DiCecco17,Bianco2019}.
R-FADs comprise thin fingertip sensor-tags (hereafter fingertip sensors)
that are interrogated by a hand-worn antenna in the 860-960 MHz band
(RFID-UHF) driven by a body-worn portable RFID reader. The feedback
to the user can be generated by a wireless companion device, like
a smartphone, a smartwatch, or earphones (Fig.~\ref{fig:Scheme-of-a}).
Numerical investigations \cite{Amendola19Numerical} demonstrated
the system is fully compliant with electromagnetic exposure limits.
A temperature-sensing R-FAD was initially exploited in \cite{Amendola19Application}
to restore the physical, thermal feeling lost because of life-saving
medical treatments and to investigate cognitive remapping of deafferented
patients. Furthermore, the R-FADs are also promising enablers for
the Tactile Internet \cite{Fanibhare21}, even though current UHF
devices still lack high data-rate low latency. Very recently \cite{Bianco2019,Bianco2020,Bianco2020bis},
dielectric sensing R-FADs exploited the automatic variation of the
impedance of the fingertip sensor's IC (integrated circuit) to sense
the dielectric constant of the touched material. This architecture
has a potential medical application as an assistive tool for visually
impaired people since the provided feedback can help them recognize
the material of the object they are handling or touching \cite{Bianco2019}.
A reliable permittivity-sensing R-FAD could be useful also for permittivity-based
industrial processes like in the food industry \cite{Routray18} and
in agriculture \cite{Magarey15,Kaptein19}. 

The practical possibility to collect well separable data when touching
the different materials is the rationale to apply a classification
procedure for the recognition of the touched object. Preliminary tests
evidenced that reader-finger communication can be critical due to
the unpredictable modality of finger-object interaction, such as the
touch pressure and the inter-user variability \cite{Bianco2020bis}.
Indeed, the specific morphology of the hand and the grasping gesture
is such that the measured data can be highly dependent on the specific
sensorized finger. Hence, single-channel measurements are not reliable.\textcolor{black}{{}
Instead, providing the hand with multiple fingertip sensors could
enable more degrees of freedom, and it is ultimately expected to improve
the reliability of the R-FAD, i.e. the capability of the device to
collect data and discriminate the dielectric constant in real use
\cite{aven2021reliability}.} \textcolor{black}{By expanding the preliminary
work in \cite{9560208}, the aim of this paper is twofold, }\textit{\textcolor{black}{i)}}\textcolor{black}{{}
quantifying the precision of the R-FAD antenna as a dielectric sensor
versus the variability of human-object interaction (both the touch-pressure
and the individual hand size and grasp modality) and }\textit{\textcolor{black}{ii)}}\textcolor{black}{{}
investigating how such variability of the measurement can be reduced
by resorting to a multi-channel configuration, wherein each finger
of the hand can be in principle independently sensorized. The information
redundancy is expected to improve the reliability of the link and
to get rid of missing readings. Experiments are carried out by using
the fingertip antenna}\textcolor{blue}{{} }\textcolor{black}{from \cite{Bianco2019},
which has been now encapsulated into soft elastomers to make it stretchable
and more comfortable for the user. }

The paper is organized as follows. The rationale of the dielectric
sensing mechanism and metrics are introduced in Section~\ref{sec:Sensing-ratio-and}.
\textcolor{black}{Section~\ref{sec:The-Multi-Channel-R-FAD} describes
the electromagnetic characterization of the multi-channel R-FAD, including
impedance matching, inter-sensors coupling, and the minimum power
the reader has to emit to have the fingertip tag responding.} Then,
Section~\ref{sec:Measurement-set-up-and} presents the measurement
procedure to determine the performance of the R-FAD as a dielectric
sensor in terms of the minimum number of collected samples needed
and the effect of touch pressure. Finally, the \textcolor{black}{reliability
of the R-FAD response} against the human variability is discussed
in Section~\ref{sec:Experimental-campaign} with the help of ten
volunteers, and the achievable improvement with the multi-channel
approach is evaluated.
\begin{figure}[t]
\begin{centering}
\includegraphics[width=8cm]{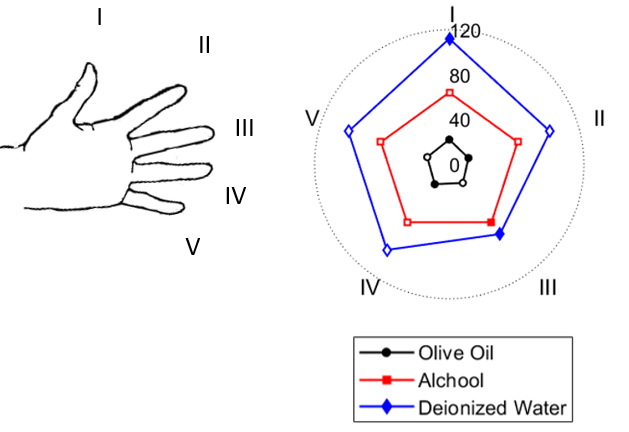}
\par\end{centering}
\caption{\textcolor{black}{Kiavat diagram visualization of digital fingerprints
corresponding to three touched dielectric materials. The unfilled
markers indicate non-read sensors.\label{fig:Kiavat-diagram-visualization-1}}}
\end{figure}
\begin{figure}[t]
\begin{centering}
\includegraphics[width=7.7cm]{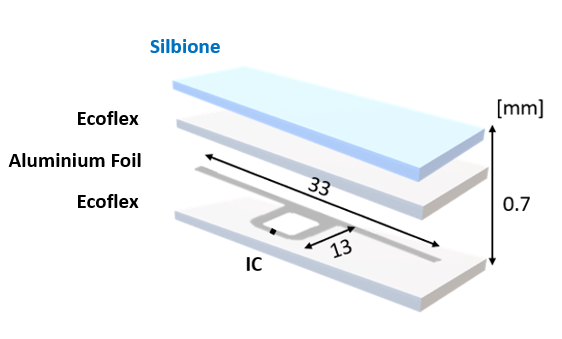}
\par\end{centering}
\caption{Exploded view of the fingertip sensor comprising a flexible antenna
over curved PET and encapsulating elastomers for comfortable application
on the finger.\label{fig:Exploded-view-of-1-1}}
\end{figure}
\begin{figure}[t]
\begin{centering}
\begin{tabular}{ccc}
\includegraphics[height=3cm]{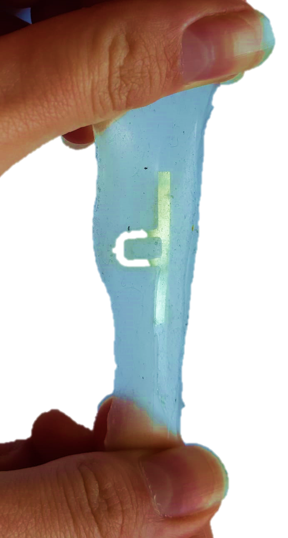} & \includegraphics[height=3cm]{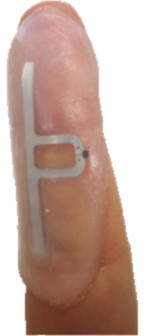} & \includegraphics[height=3cm]{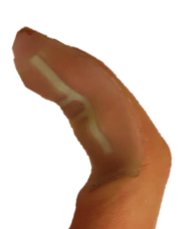}\tabularnewline
(a) & (b) & \enskip{}(c)\tabularnewline
\end{tabular}
\par\end{centering}
\caption{Final prototypes of fingertip sensor: (a) stretching, (b) application
on a fingertip, and (c) finger bending. \label{fig:Fingersensor-prototype.-(a)-1-1-1-1}}
\end{figure}

\section{Rationale of Dielectric Sensing and Related Metrics\label{sec:Sensing-ratio-and}}

\subsection{Dielectric Sensing by Auto-tuning RFID ICs}

Dielectric sensing by R-FADs is achieved by exploiting the unique
feature of auto-tuning ICs, that can modify their internal radiofrequency
impedance to compensate for possible mismatching with the antenna
and to maximize the power that the harvesting antenna delivers to
the IC itself. Auto-tuning ICs can return a digital metric, denoted
as sensor code ($s$), which is an integer number that is\textbf{
}proportional to the retuning effort \cite{Caccami18ter}. The equivalent
circuit of the microchip can be modeled as a resistor in parallel
with a ladder of capacitors with overall capacitance $C_{IC}$ spanning
from a minimum $C_{min}$ to a maximum value with incremental step
$C_{S}$ so that 

\begin{equation}
C_{IC}\left(s\right)=C_{min}+sC_{S}.\label{eq:capacitance}
\end{equation}
The overall variable susceptance of the IC is derived by the auto-tuning
condition \cite{Caccami18ter}

\begin{equation}
\left|B_{IC}\left(s\right)+B_{A}\left(\epsilon\right)\right|=0\label{eq:self-tuning_equation}
\end{equation}
where $B_{IC}\left(s\right)=2\pi fC_{IC}\left(s\right)$ and $B_{A}\left(\epsilon\right)$
are the susceptances of the IC and the antenna, respectively, $\epsilon$
is the dielectric constant of the touched object, and $f$ is the
frequency.

Equations (\ref{eq:capacitance}) and (\ref{eq:self-tuning_equation})
hold in a linear range $S_{min}\leq s\leq S_{max}$, outside which
saturation occurs \cite{Bianco2020}. In the linear range, the index
$s$ can be related to the antenna input parameters by inverting (\ref{eq:self-tuning_equation})
and accounting for the saturation \cite{Bianco21}

\begin{equation}
s\left(\epsilon\right)=\textnormal{nint}\left\{ -\frac{1}{C_{S}}\left[C_{IC}\left(S_{min}\right)+\frac{B_{A}\left(\epsilon\right)}{2\pi f}\right]\right\} .
\end{equation}
Possible baselines caused by the specific manufacturing of the fingertip
sensor can be removed by employing a \emph{differential sensor code
}($\Delta s$) \cite{Bianco2020bis} 
\begin{equation}
\Delta s\left(\epsilon\right)=s\left(1\right)-s\left(\epsilon\right)\label{eq:differential_SC}
\end{equation}
 where $s\left(1\right)$ is the calibration value obtained when the
hand is not touching anything ($\epsilon=1$).

\subsection{Sensing R-FAD Metrics}

\begin{figure}[t]
\begin{centering}
\includegraphics[width=6.8cm]{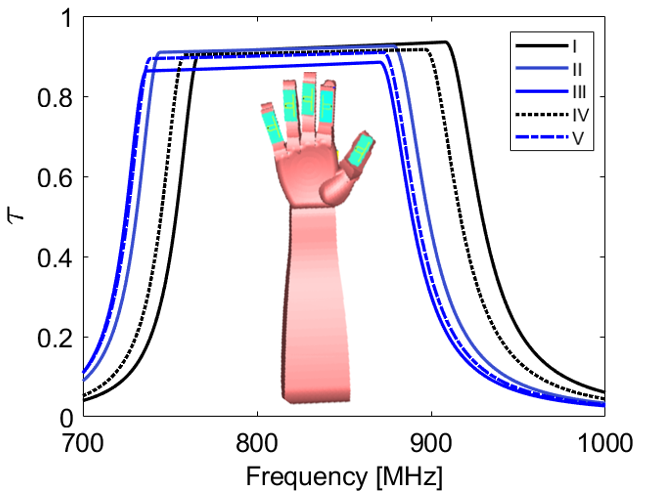}
\par\end{centering}
\caption{The simulated power transmission coefficient of each fingertip sensor.
In the inset: the simplified homogeneous model of the hand from \cite{Bianco2020}.
\label{fig:Simulated-power-transmission-1}}
\end{figure}
\begin{figure}[t]
\begin{centering}
\includegraphics[width=8cm]{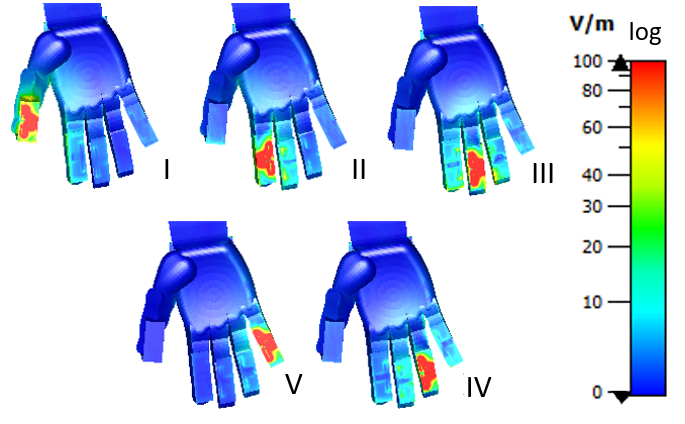}
\par\end{centering}
\caption{Numerical simulations of the inter-sensor electromagnetic coupling
of the multi-channel R-FAD. Map of the E-field when only one of the
five ports is sourced by 1 W power. Roman numerals identify the finger
of the active antenna. \label{fig:Numerical-simulations-of-1-1}}
\end{figure}
 The information provided by a multi-channel R-FAD is hereafter exploited
by introducing the following sensing metrics: the \textit{digital
fingerprint} and the \textit{averaged fingerprint.}

\subsubsection{Digital fingerprint}

Multi-channel R-FADs comprise $1<N\leq5$ fingertip sensors. When
the wearer touches a dielectric material of permittivity $\epsilon_{m}$,
$N$ sensor codes $\left\{ s_{n}(\epsilon_{m})\right\} _{n=1,...,N}$
are returned, and the set of differential sensor codes $\left\{ \Delta s_{n}(\epsilon_{m})\right\} _{n=1,...,N}$
composes the\textit{ }digital fingerprint\emph{ }

\begin{equation}
\mathbf{F}\left(\epsilon_{m}\right)=\left\{ \triangle s_{1}\left(\epsilon_{m}\right),...,\triangle s_{N}\left(\epsilon_{m}\right)\right\} .\label{eq:fingerprint-1}
\end{equation}
The $N$-dimensional fingerprint can be visualized through Kiviat
diagrams (Fig.~\ref{fig:Kiavat-diagram-visualization-1}). If the
\emph{k}th sensor is not read by the reader in the specific condition,
its differential sensor code $\triangle s_{k}$ is replaced by the
mean over the other responsive fingertip sensors so that its differential
sensor code is computed as

\begin{equation}
\triangle s_{k}\left(\epsilon_{m}\right)=\frac{1}{N_{n}}\sum_{i=1}^{N_{n}}\triangle s_{i}(\epsilon_{m}),\label{eq:averageSC-1}
\end{equation}
where $N_{n}$ is the number of responsive fingertip sensors\textit{.}

\subsubsection{Averaged fingerprint}

The mean of the differential sensor codes of all sensors of the hand
gives the averaged fingerprint
\begin{center}
\begin{equation}
\overline{F}\left(\epsilon_{m}\right)=\frac{1}{5}\sum_{i=1}^{5}\triangle s_{i}(\epsilon_{m}).\label{eq:averaged fingerprint}
\end{equation}
\par\end{center}

\section{\textcolor{black}{The Multi-Channel R-FAD} \label{sec:The-Multi-Channel-R-FAD}}

\begin{table}[t]
\caption{Electrical characteristics of the employed materials ($f=867$ MHz).
\label{tab:Electric-characteristics-of-1-1-1}}

\centering{}%
\begin{tabular}{l|l|l}
\textbf{Material} & \textbf{$\mathbf{\sigma}\ \left[\nicefrac{\mathbf{S}}{\textnormal{\ensuremath{\mathbf{m}}}}\right]$} & \textbf{$\mathbf{\epsilon}$}\tabularnewline
\hline 
\hline 
$\textnormal{Ecoflex}^{\textnormal{TM}}$ 00-30 & $\sigma=0.007$ & $2.7$\tabularnewline
\hline 
$\textnormal{Silbione}^{\textnormal{TM}}$ & $\sigma=0.012$ & $2.5$\tabularnewline
\hline 
Closed-cell PVC foam \cite{Lodato14} & $\sigma=2.2\cdot10^{-5}$ & $2.3$\tabularnewline
\hline 
Homogeneous body tissue \cite{Amendola19Numerical} & $\sigma=0.62$ & $30$\tabularnewline
\hline 
Olive oil \cite{Bianco2020bis} & $\sigma=0.026$ & $3$\tabularnewline
\hline 
Ethyl alcohol $\left(90\%\right)$ \cite{Sengwa03} & $\sigma=10^{-5}$ & $17$\tabularnewline
\hline 
Deionized water \cite{Malmberg56} & $\sigma=0.05$ & $78$\tabularnewline
\end{tabular}
\end{table}
 The layout of the reference finger antenna to evaluate the above-defined
metrics is derived from \cite{Bianco2019}. It is a T-matched aluminum
dipole soldered to Axzon Magnus-S3 ICs having input admittance $G_{IC}=0.482$
mS and equivalent capacitance derived from (\ref{eq:capacitance})
with parameters $C_{min}=1.9$ pF, $C_{S}=3.1$ fF, $\{S_{min},S_{max}\}=\{80,400\}$. 

The dipole is encapsulated between two layers of $\textnormal{Ecoflex}^{\textnormal{TM}}$
$00$-$30$ (by Smooth-On; thickness $\sim0.3$ mm each) silicone,
which are cured together with a layer of $\textnormal{Silbione}^{\textnormal{TM}}$
(by Elkem Silicones; thickness $\sim0.1$ mm; Fig.~\ref{fig:Exploded-view-of-1-1}).
The $\textnormal{Ecoflex}^{\textnormal{TM}}$ acts as a coating film
for the sensor, whereas the $\textnormal{Silbione}^{\textnormal{TM}}$
ensures adherence to the finger's skin. The electrical properties
of the two silicone rubbers, as measured with a two-port ring resonator
\cite{Bernard91}, are listed in Table~\ref{tab:Electric-characteristics-of-1-1-1}.
\textcolor{black}{The resulting fingertip sensors are highly flexible,
stretchable, biocompatible, and can be removed and re-applied comfortably
(Fig.~\ref{fig:Fingersensor-prototype.-(a)-1-1-1-1}). They also
withstand disinfection by hydrogen peroxide. The reader antenna is
a folded patch on a closed-cell PVC (polyvinyl chloride) foam as in
\cite{Amendola19Numerical} that is placed on the back of the hand.}

\subsection{Impedance Matching }

The fingertip sensor is numerica\textcolor{black}{lly simulated}\footnote{\textcolor{black}{Numerical simulations by CST - Microwave Studio
2018.}}\textcolor{black}{{} when it is placed on a homogeneous phantom of the
hand (from \cite{Bianco2020}). Fig. \ref{fig:Simulated-power-transmission-1}
shows the power transfer coefficients $\tau\left(f\right)$, which
is the fraction of the RF power that is harvested by the antenna and
then transferred to the IC. The power transfer coefficient is computed
by accounting for the auto-tuning according to (\ref{eq:self-tuning_equation})
as in \cite{Bianco2019},}

\textcolor{black}{
\begin{equation}
\tau\left(f\right)=\begin{cases}
\frac{4G_{IC}G_{A}}{\left|G_{IC}+G_{A}+j\left(B_{A}+2\pi fC_{min}\right)\right|^{2}}\\
\textnormal{if}\ B_{A}>-2\pi fC_{min}\\
\frac{4G_{IC}G_{A}}{\left(G_{IC}+G_{A}\right)^{2}}\\
\textnormal{elsewhere}\\
\frac{4G_{IC}G_{A}}{\left|G_{IC}+G_{A}+j\left[B_{A}+2\pi f\left(C_{min}+sC_{S}\right)\right]\right|^{2}}\\
\textnormal{if}\ B_{A}<-2\pi f\left(C_{min}+S_{max}C_{S}\right)
\end{cases}
\end{equation}
where $G_{A}$ is the conductance seen by the IC. Thanks to the auto-tune
property of the IC, \ensuremath{\tau} exhibits the typical nearly
flat profile over broadband. The power transfer coefficient at the
EU RFID frequency is }$\tau\left(867\ \textnormal{MHz}\right)>0.85$
for any fingertip sensor.

\subsection{Inter-antenna Coupling}

The electromagnetic coupling among fingertip antennas when all the
fingers are sensorized is here evaluated. Coupling is relevant to
the cross-sensitivity of multiple sensors on the same hand, namely
the possible dependence of the response of any considered fingertip
sensor on the others. 

The electric near field is computed over the fingers when each antenna
is singularly excited while the others are not. The color map in Fig.~\ref{fig:Numerical-simulations-of-1-1}
shows that, by the naturally decoupling effect of the human tissue
absorption, the strength of the fields on non-excited fingertip sensors
is less than 10\% that of the excited one. The coupling can be quantified
through Kurokawa's generalized scattering matrix $\mathbf{\mathbf{K}}$
accounting for the ICs complex impedances \cite{Nikitin05,Kurokawa65}

\begin{equation}
\mathbf{K}=\mathbf{G}\left(\mathbf{Z}-\mathbf{H^{+}}\right)\left(\mathbf{Z+\mathbf{H}}\right)^{-1}\mathbf{G}^{-1}.
\end{equation}
$\mathbf{Z}$ is the antennas' impedance matrix, ``$^{+}$'' indicates
the complex conjugate transpose matrix, and $\mathbf{G}$ and $\mathbf{H}$
are diagonal matrices whose non-null components are equal to $0.5\left[\textnormal{Re}\left(Z_{C}\right)\right]^{-0.5}$
and $Z_{C}$, respectively. $Z_{C}$ is the complex impedance of the
ICs assumed fixed in the middle range ($Z_{C}=\left(2.8-j76\right)\ \Omega$
\cite{Bianco21}). The normalized scattering matrix of the fingertip
sensors' ports at $f=867$ MHz is 
\begin{table}[H]
\[
\frac{\left|\mathbf{K}_{jk}\right|}{\underset{j,k}{\max}\left[\left|\mathbf{K}_{jk}\right|\right]}=
\]
\begin{equation}
=\frac{1}{100}\left[\begin{array}{ccccc}
\mathbf{98.74} & 0.15 & 0.09 & 0.04 & 0.01\\
0.15 & \mathbf{88.71} & 2.89 & 0.43 & 0.05\\
0.09 & 2.89 & \mathbf{91.38} & 0.86 & 0.01\\
0.04 & 0.43 & 0.86 & \mathbf{97} & 0.08\\
0.01 & 0.05 & 0.53 & 0.08 & \mathbf{100}
\end{array}\right].
\end{equation}
\end{table}

The inter-sensor coupling is rather modest sinc\textcolor{black}{e
the off-diagonal elements are $\lesssim$ 3\% of the diagonal ones.
Further numerical simulations assessed that there is no significant
change in the electromagnetic coupling even when the fingers are bent
during the grasping gesture. Accordingly, the six-port multi-channel
R-FAD can be reduced to five two-port models, one for each reader-tag
link \cite{Bianco2020}. }

\textcolor{black}{Therefore the mutual hand-finger links are mostly
unaffected by the simultaneous wearing of multiple se}nsors and, consequently,
the cross-sensitivity of the sensors is negligible.

\subsection{Electromagnetic Performance}

\begin{figure}[t]
\begin{centering}
\textcolor{black}{}%
\begin{tabular}{cc}
\textcolor{black}{\includegraphics[height=4cm]{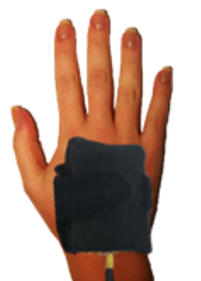}} & \textcolor{black}{\includegraphics[height=4cm]{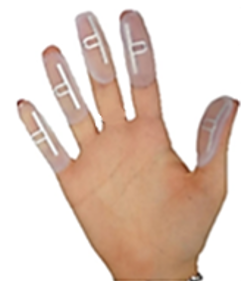}}\tabularnewline
\textcolor{black}{\enskip{}(a)} & \textcolor{black}{\enskip{}\,\enskip{}\enskip{}(b)}\tabularnewline
\end{tabular}
\par\end{centering}
\textcolor{black}{\caption{(a) Back-hand reader antenna. (b) The five fingertip sensors worn
simultaneously. \label{fig:Layouts-of-R-FAD-1-1-1}}
}
\end{figure}

The electromagnetic performance of the prototypes is characterized
by the Voyantic Tagformance Pro station. \textcolor{black}{The reader
antenna is positioned on the back of the hand (Fig.~\ref{fig:Layouts-of-R-FAD-1-1-1}(a)),
and it is moderately conformable to the user's hand morphology.} The
measurements are made by asking a volunteer to first wear just one
sensor at a time on each finger (single-channel) and then all the
five sensors (multi-channel; Fig.~\ref{fig:Layouts-of-R-FAD-1-1-1}(b))
with the hand straight open without touching\textcolor{black}{{} anything.
The orientation of the fingertip sensors was experimentally determined
to minimize the body shadowing caused by the nearby fingers. In all
cases, the Tagformance raises the emitted power until the }\textit{\textcolor{black}{k}}\textcolor{black}{th
fingertip sensor starts responding, and that threshold value is the
}\textit{\textcolor{black}{turn-on power}}\textcolor{black}{{} of that
fingertip antenna in the current condition. The turn-on power was
predicted by means of the transducer power gain and the IC sensitivity
\cite{Bianco2020} and is the considered communication performance
metric. }

\begin{figure}[t]
\begin{centering}
\textcolor{black}{\includegraphics[width=6.7cm]{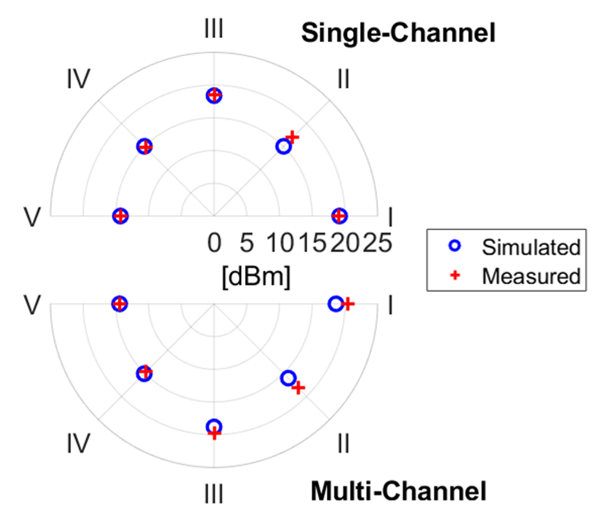}}
\par\end{centering}
\textcolor{black}{\caption{Simulated and measured turn-on powers (at $f=867$ MHz) in the case
of the single-channel (only a sensorized fingertip at a time) and
the multi-channel (all fingers simultaneously sensorized) on-body
link. \label{fig:Turn-on-powers-measured-1}}
}
\end{figure}

\textcolor{black}{As expected by the previous numerical analysis,
the recorded turn-on powers confirm no significant difference in the
case of single-sensor and multi-sensors configurations (Fig.~\ref{fig:Turn-on-powers-measured-1})
since the antennas are negligibly coupled. All the fingertip sensors
require the generator to feed the reader antenna with less than $25$
dBm, compatible with hand-held readers having a size similar to watches
or key-fobs. For this input power, numerical simulations (not reported
here) returned a maximum SAR (Specific Absorption Rate) in the hand
of about 2 $\nicefrac{W}{kg}$ in the most conservative case of continuous
emission. This value is fully compliant with the exposure limit (4
$\nicefrac{W}{kg}$) \cite{recommendation1999limitation}.}

\section{Performance as Dielectric Tactile Sensor \label{sec:Measurement-set-up-and}}

This Section introduces the experimental set-up and the processing
to evaluate the performance of the R-FAD as a dielectric sensor according
to the metrics introduced in Section II. In particular, this Section
discusses the minimum number of se\textcolor{black}{nsor code samples
that guarantees a reliable measurement outcome, and it investigates
the sources of uncertainty posed by the touch pressure variability
and different containers of the liquids. }
\textcolor{black}{}
\begin{figure}[t]
\begin{centering}
\textcolor{black}{}%
\begin{tabular}{c}
\textcolor{black}{\includegraphics[height=4.6cm]{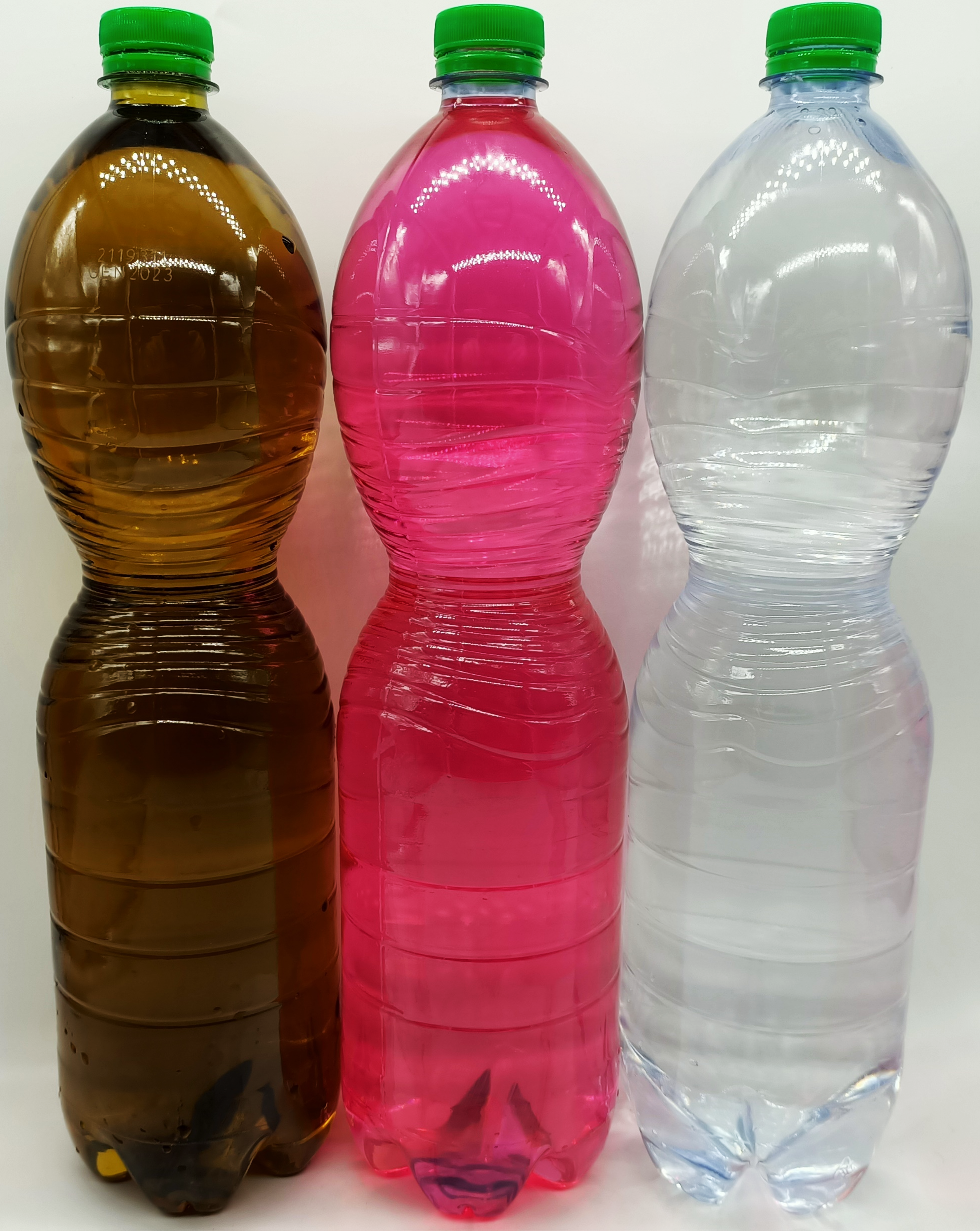}}\tabularnewline
\textcolor{black}{(a)}\tabularnewline
\end{tabular}\textcolor{black}{}%
\begin{tabular}{c}
\textcolor{black}{\includegraphics[width=1.5cm]{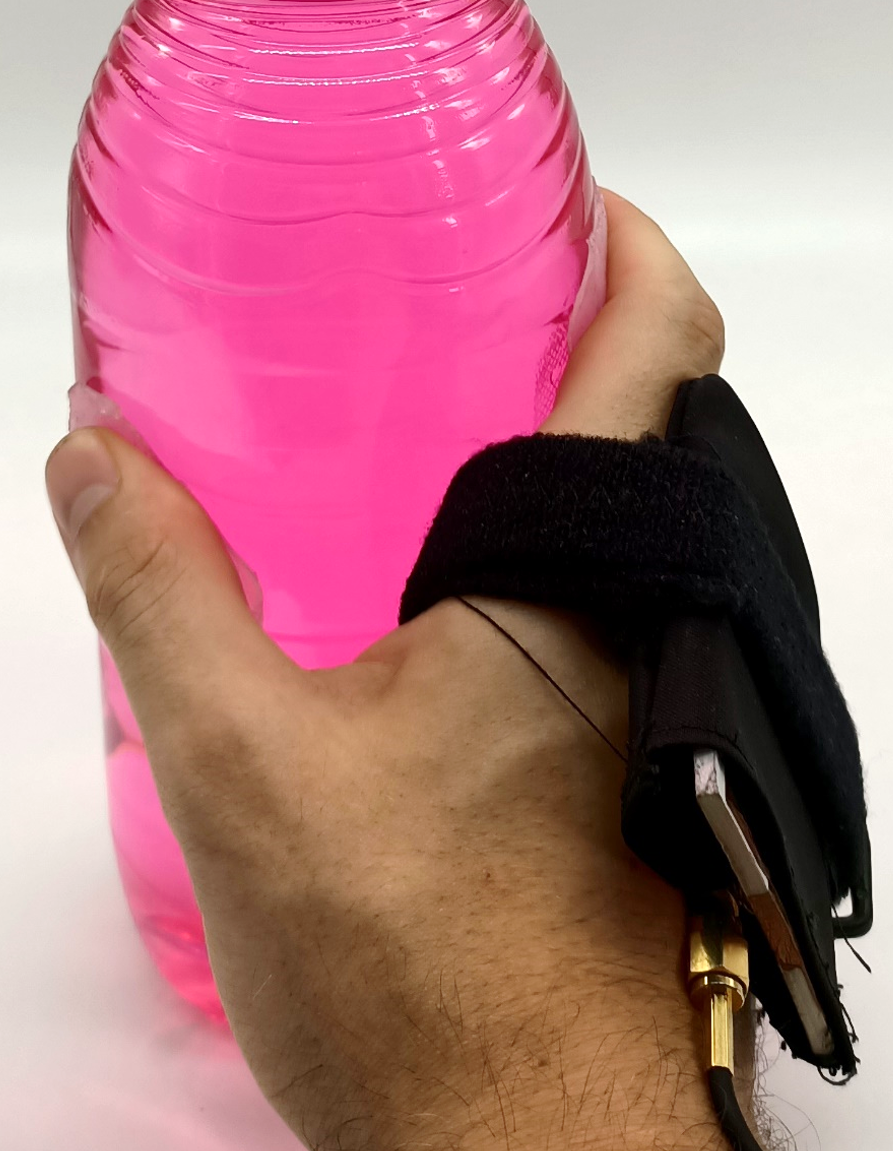}}\tabularnewline
\textcolor{black}{(b)}\tabularnewline
\textcolor{black}{\includegraphics[width=1.5cm]{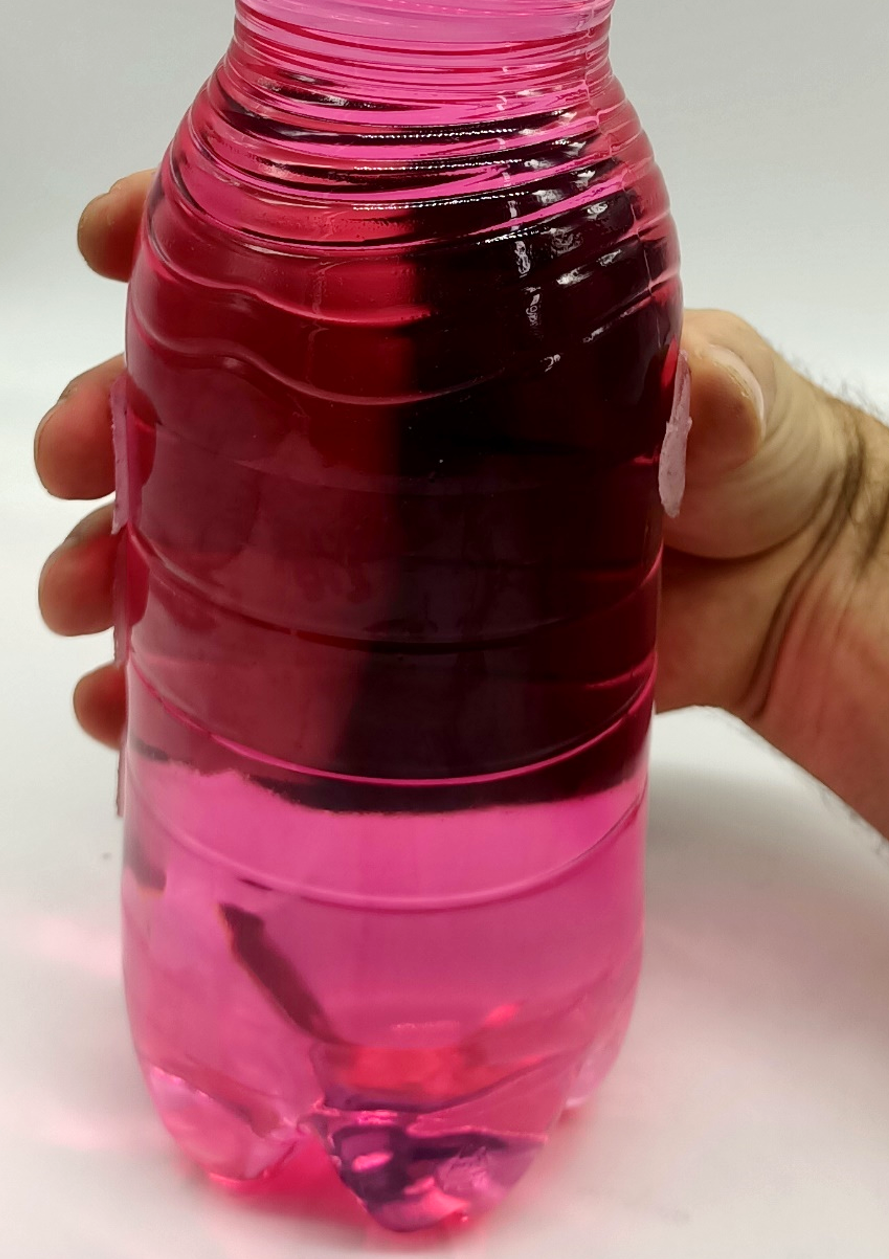}}\tabularnewline
\textcolor{black}{(c)}\tabularnewline
\end{tabular}
\par\end{centering}
\textcolor{black}{\caption{Experimental set-up. (a) Three PET bottles filled with the three dielectric
liquids. (b,c) The performed reference grasp gesture to touch the
bottles. \label{fig:Three-bottles-filled-1}}
}
\end{figure}
\textcolor{black}{}
\begin{figure}[t]
\begin{centering}
\textcolor{black}{\includegraphics[width=8.8cm]{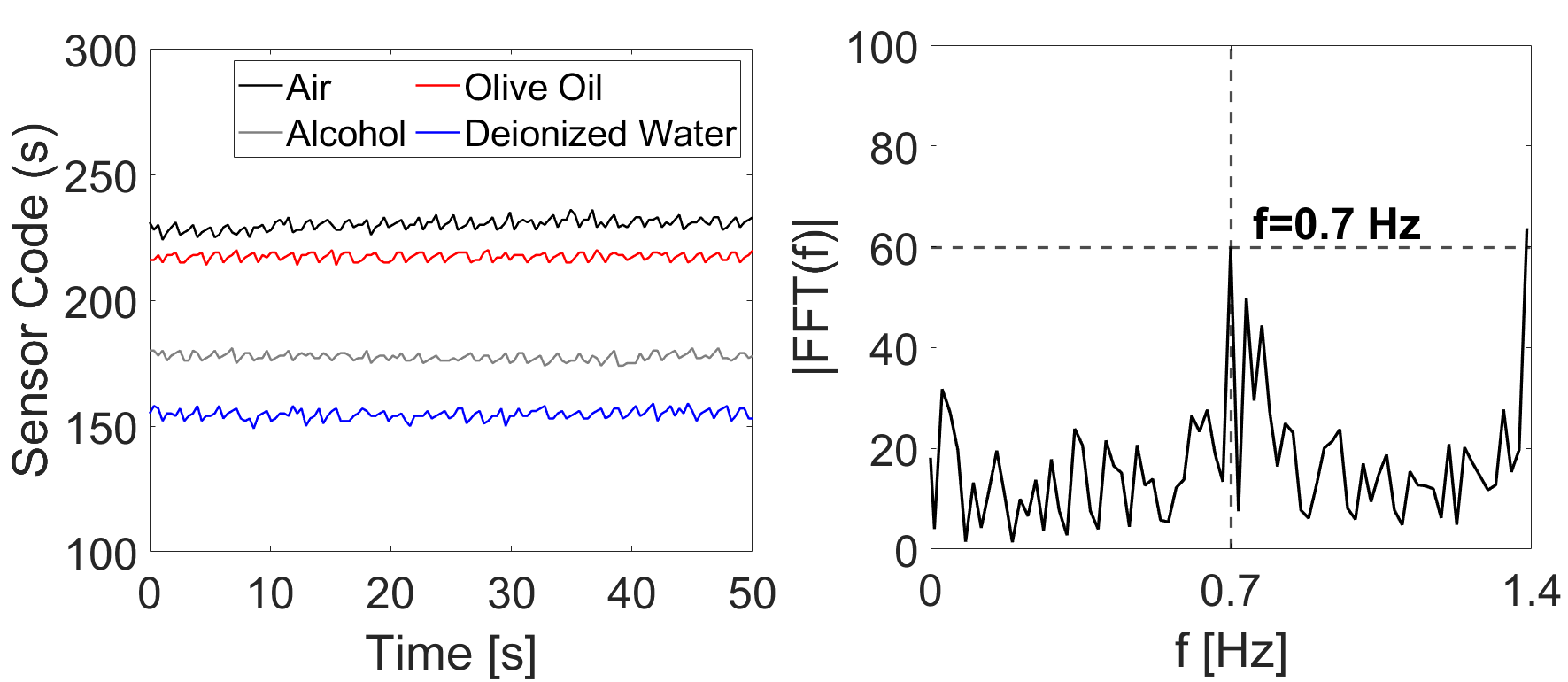}}
\par\end{centering}
\begin{centering}
\textcolor{black}{}%
\begin{tabular}{cc}
\textcolor{black}{\hspace{1cm}(a)\hspace{3.4cm}} & \textcolor{black}{(b)}\tabularnewline
\end{tabular}
\par\end{centering}
\textcolor{black}{\caption{\textcolor{black}{(a) Continuous acquisition of the sensor code $s$
while the bare sensor (not attached to the finger) is in the air and
when it adheres to a bottle filled with a given liquid. (b) Amplitude
spectrum of $s(t)$ evaluated through Fast Fourier Transform (FFT).}
\label{fig:SC-fluctuations.-(a)-1}}
}
\end{figure}
\textcolor{red}{}
\begin{figure}[t]
\begin{centering}
\includegraphics[width=8.8cm]{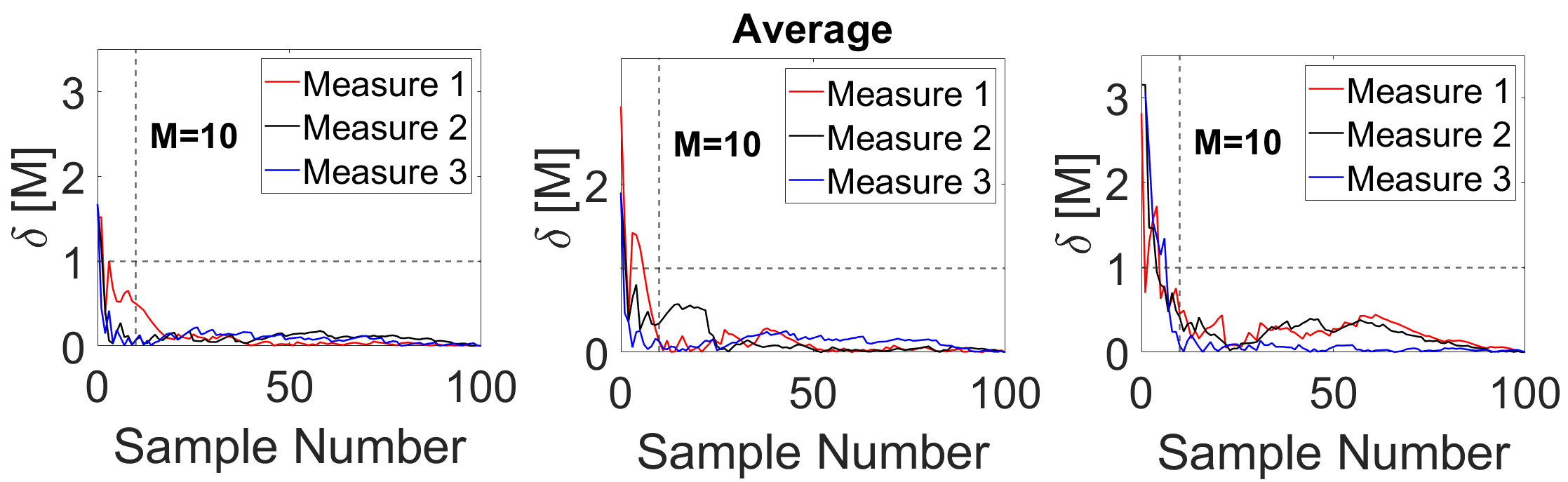}
\par\end{centering}
\begin{centering}
\begin{tabular}{ccc}
\hspace{0.3cm}\hspace{0.3cm}(a)\hspace{2.2cm} & (b) & \hspace{2.2cm}(c)\tabularnewline
\end{tabular}
\par\end{centering}
\begin{centering}
\includegraphics[width=8.8cm]{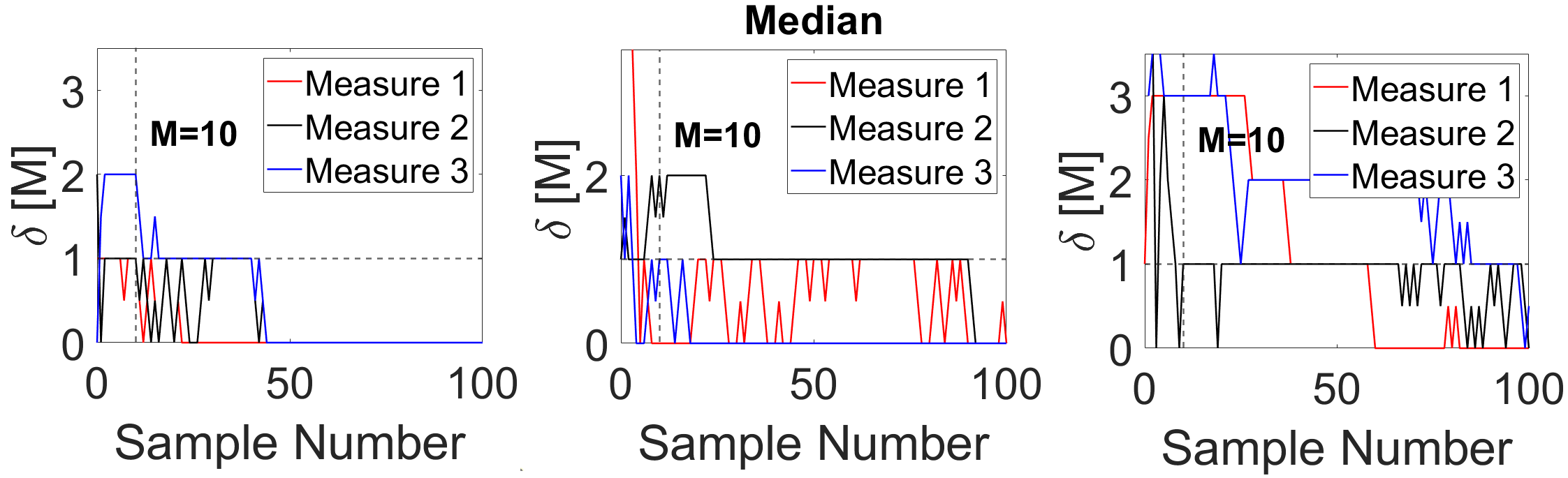}
\par\end{centering}
\begin{centering}
\begin{tabular}{ccc}
\hspace{0.3cm}\hspace{0.3cm}(d)\hspace{2.2cm} & (e) & \hspace{2.2cm}(f)\tabularnewline
\end{tabular}
\par\end{centering}
\textcolor{red}{\caption{\textcolor{black}{Convergence error vs the permanent regime over an
increasing number of samples when the sensorized finger touches three
different materials: olive oil, alcohol, and deionized water. (a,b,c)
Convergence error of the average. (d,e,f) Convergence error of the
median.\label{fig:Stability-of-a-1}}}
}
\end{figure}
\begin{figure}[t]
\begin{centering}
\includegraphics[width=7.5cm]{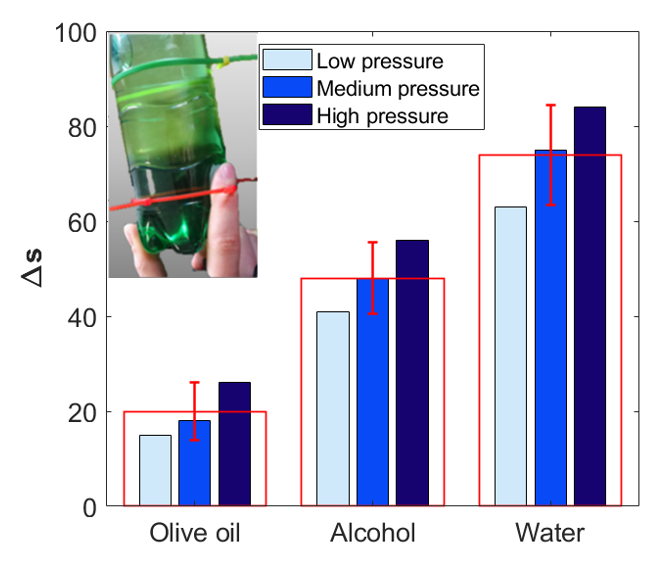}
\par\end{centering}
\caption{Measured digital fingerprints corresponding to increasing touch pressures
on the bottles filled by the three liquids. The wide red bars indicate
the mean values, and the standard deviation $\sigma_{p}$ is also
represented. \label{fig:Digital-fingerprints-when}}
\end{figure}
\begin{figure}[t]
\begin{raggedright}
\includegraphics[width=8.9cm]{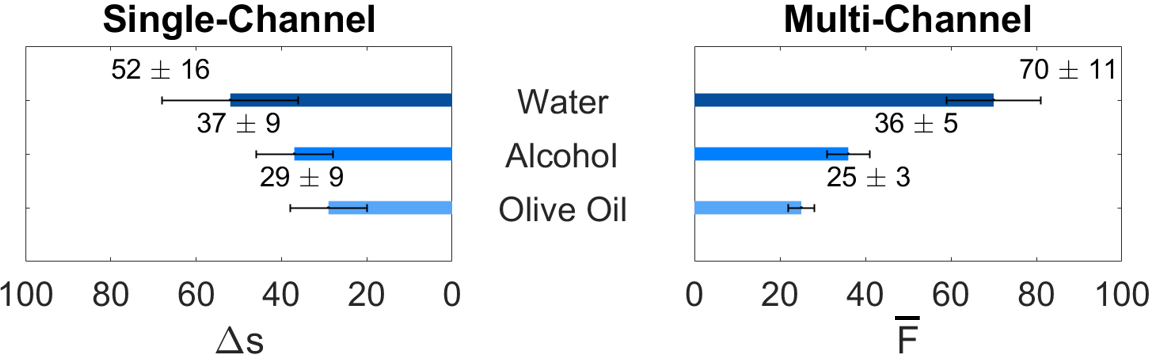}
\par\end{raggedright}
\caption{Comparison between \textcolor{black}{the differential sensor code
$\triangle s$ }of the single-channel (fingertip sensor II) and the
averaged fingerprint $\overline{F}$ of the multi-channel R-FAD. Black
segments indicate uncertainties. \label{fig:Digital-contrasts-of-1}}
\end{figure}
\begin{figure}[t]
\begin{centering}
\includegraphics[width=8.8cm]{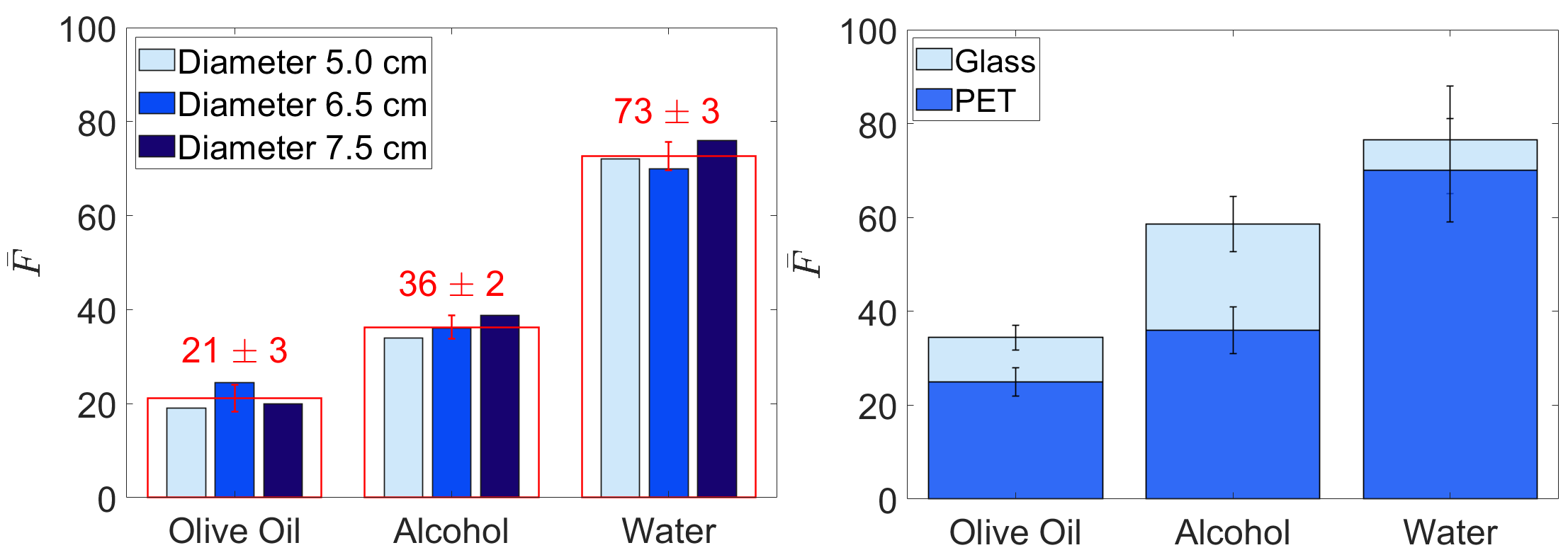}
\par\end{centering}
\begin{centering}
\begin{tabular}{cc}
\hspace{1cm}(a)\hspace{3.4cm} & (b)\tabularnewline
\end{tabular}
\par\end{centering}
\centering{}\caption{\textcolor{black}{Averaged fingerprints $\overline{F}$ of the multi-channel
R-FAD when (a) touching PET bottles with different diameters and (b)
touching bottles glass in comparison with the PET case. \label{fig:Averaged-fingerprints-of}}}
\end{figure}
\begin{figure}[t]
\begin{centering}
\includegraphics[width=8cm]{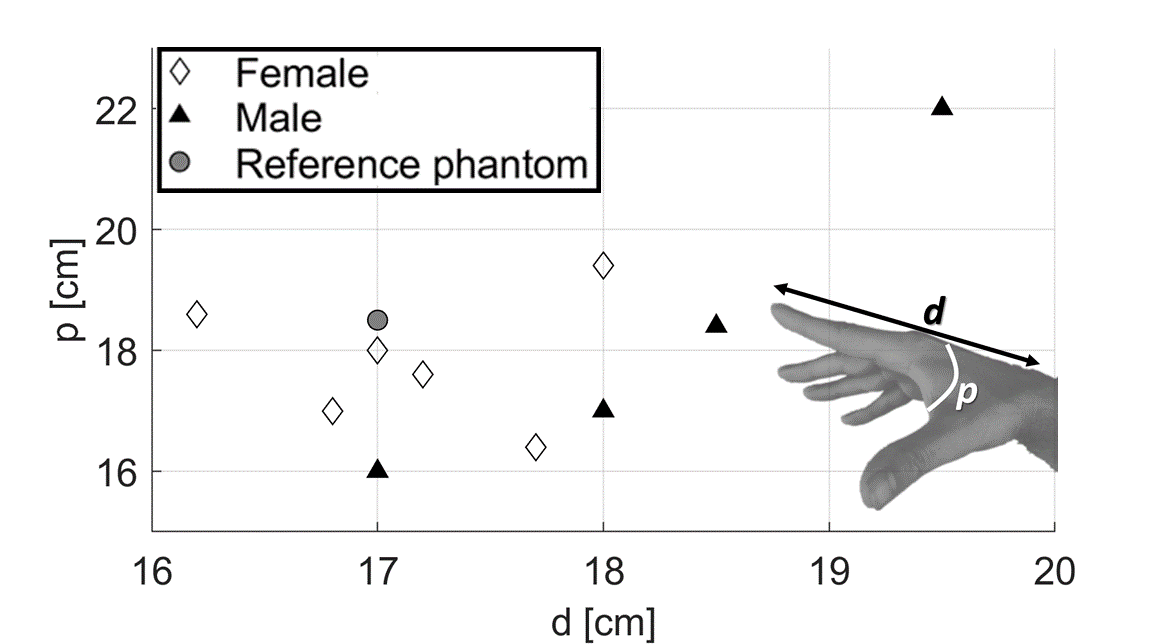}
\par\end{centering}
\caption{Hand sizes (wrist-fingertip distance \emph{d} and palm perimeter \emph{p})
of the ten volunteers. \label{fig:Test-population-hand-1-1}}
\end{figure}
\begin{figure}[t]
\begin{centering}
\includegraphics[height=4cm]{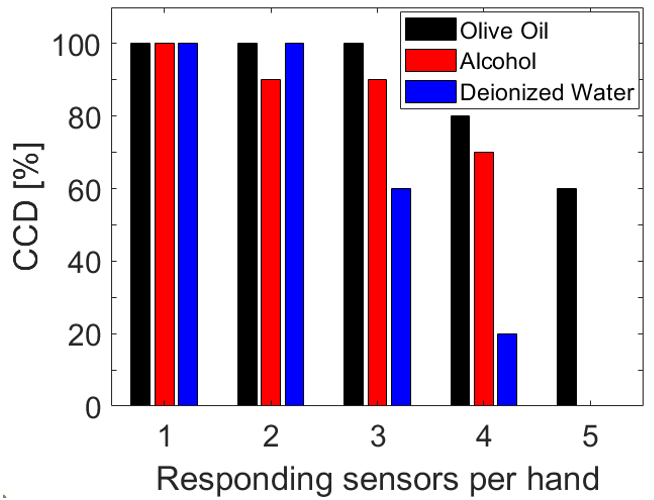}\includegraphics[height=4cm]{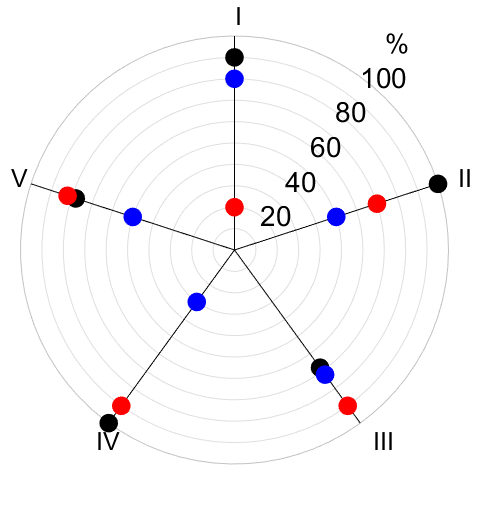}
\par\end{centering}
\begin{centering}
(a) $\qquad$$\qquad$$\qquad$$\qquad$ $\qquad$ (b)
\par\end{centering}
\caption{(a) Complementary cumulative distribution (CCD) bar of the simultaneously
responding finger-sensors of the same hand over the test population.
(b) Percentages of the readable sensors per specific finger of the
same hand of the volunteers when touching the three objects (color
markers as in the inset on the left). \label{fig:Bar-plot-over}}
\end{figure}
\begin{figure*}[t]
\begin{centering}
\includegraphics[width=17cm]{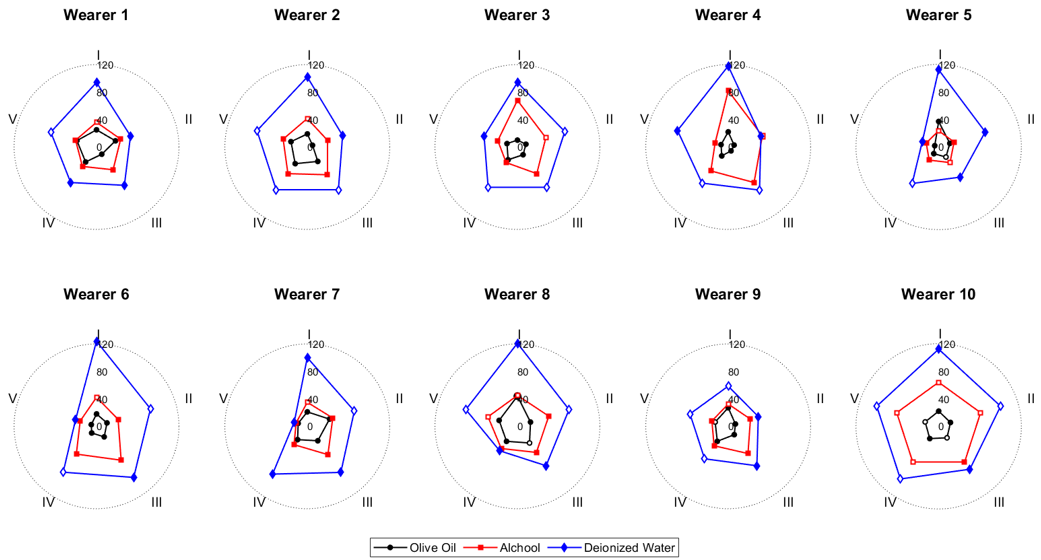}
\par\end{centering}
\centering{}\caption{Digital fingerprints of the test population when the multi-channel
R-FADs are naturally touching the three materials with no control
on the pressure. The unfilled markers symbolize non-read sensors.
\label{fig:Digital-fingerprint-of}}
\end{figure*}
\begin{figure}[t]
\begin{raggedright}
\includegraphics[width=8.7cm]{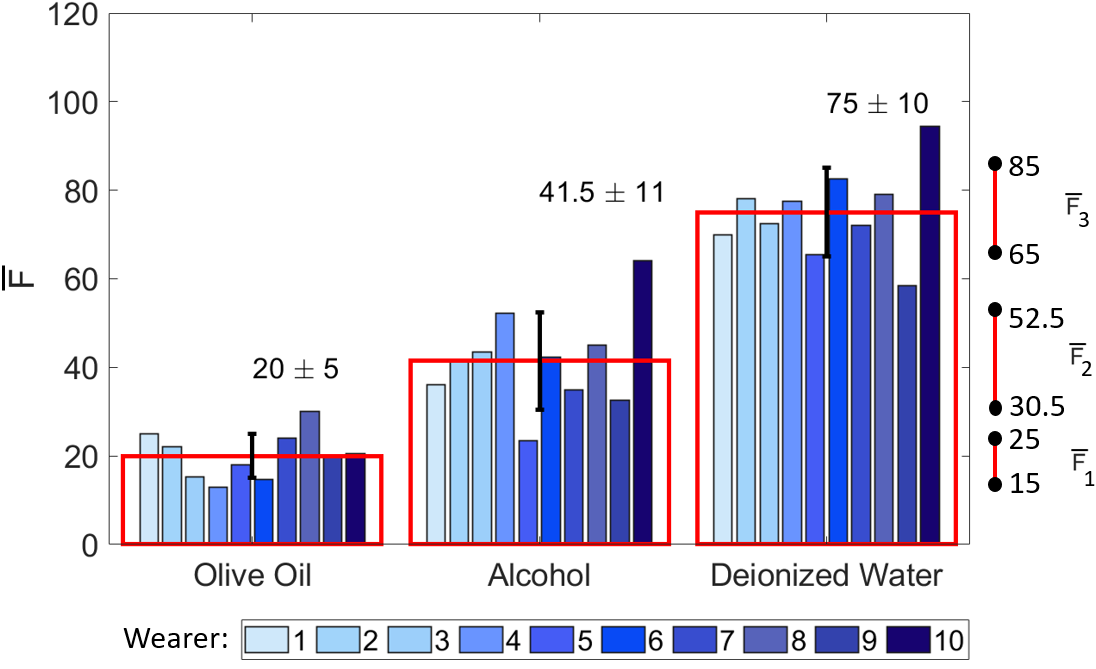}
\par\end{raggedright}
\caption{Averaged fingerprints $\overline{F}$ of the ten volunteers when touching
the three objects with increasing permittivity. The macro red bars
indicate the average and standard deviation over the population. On
the right, the overall span of the $\overline{F}\pm\sigma$ for the
three materials. \label{fig:Averaged-fingerprints-}}
\end{figure}

\subsection{Reference Objects\label{subsec:Materials}}

The experimental reference set-up compris\textcolor{black}{es three
equal-size PET (polyethylene terephthalate, $\epsilon=2.5$ \cite{Bianco2020})
bottles having a diameter 6.5 cm and thickness 0.3 mm. Different containers
from the reference one are utilized to assess their impact on sensing
in Section \ref{subsec:Container-Variation}. The containers are filled
with three dielectric liquids of common use: olive oil, ethyl alcohol,
and deionized water (Fig.~\ref{fig:Three-bottles-filled-1}(a); electric
proper}ties in Table~\ref{tab:Electric-characteristics-of-1-1-1}).
The liquids have high dielectric contrast, so they are suitable to
test the sensors' response in both low-medium and medium-high permittivity
ranges. The R-FAD wearer is asked to normally grasp t\textcolor{black}{he
bottles (}Fig.~\ref{fig:Three-bottles-filled-1}(b,c)) and the sensor
code is collected. 

\subsection{Measurement Time Window}

\textcolor{black}{When the sensor contacts an object,} the auto-tune
functionality of the microchip activates and after an initial transient,
it stabilizes, fluctuating around a baseline that depends on the touched
material. The fluctuation is the consequence of the charge and discharge
of the internal equivalent capacitor during the RFID interrogation.
\textcolor{black}{Accordingly, the measurement time window must be
large enough to fade the touch transient and then to drop out the
fluctuations by averaging. The typical behavior of the fluctuation
in the static condition, namely when just the bare sensor was directly
attached to the bottle without the presence of the finger, is a sawtooth
profile (Fig.~\ref{fig:SC-fluctuations.-(a)-1}(a)). Independently
from the contacted material, it is characterized by a dominant frequency
of $0.7$~Hz (Fig.~\ref{fig:SC-fluctuations.-(a)-1}(b)). Hence,
at least one period ($1.4$~s) must be sampled to remove fluctuations
by averaging. }

\textcolor{black}{The duration of the transient when the wearer touches
the bottle is estimated through the evaluation of the convergence
error $\delta$. The convergence error is used to derive the number
of samples needed to obtain stable metrics and is d}efined as \cite{Camera21}

\begin{equation}
\delta[M]=\sigma_{s}[M]-\sigma_{s}[M_{\infty}].
\end{equation}
 $\sigma_{s}[M]$ is the standard deviation of the sensor code evaluated
on $M$ samples, whereas $\sigma_{s}[M_{\infty}]$ approximates the
asymptotic value corresponding to $M_{\infty}=100$ samples. The minimum
number $M_{0}$ of samp\textcolor{black}{les for averaging to return
a stable measurement is determined so that $\delta[M_{0}]<1$. This
analysis also accounts for the effect of fluctuations. $M_{0}$ was
evaluated through measurements of the sensor code when the fingertip
sensors touched three times each liquid. Fig.~\ref{fig:Stability-of-a-1}(a,b,c)
shows that $M_{0}$ is higher as the material's permittivity increases
so that the value $M_{0}=10$ is considered as a valid trade-off for
all three materials. Since the sampling period of the RFID reader
is $0.7$~s, the acquisition time window will be therefore $7$~s,
which is also suitable to drop out the natural fluctuation of the
sensor code, as described above. Another option for data analysis
could be the median that is more robust against outliers. However,
the time required to have a stable measurement is longer than the
mean (Fig.~\ref{fig:Stability-of-a-1}(d,e,f)), taking up to $90$
samples versus $10$ samples for the average to return a convergence
error lower than a single unit. Accordingly, the mean operator will
be hereafter used. Measurements could be significantly sped up by
accepting a convergence error of $2$ units instead of $1$, so to
add one unit of uncertainty. In this case, the measurement time could
be reduced to just $2$ seconds. This choice is still compliant with
the possibility to discriminate the considered liquids. To be more
conservative, the convergence error lower than $1$ unit and the corresponding
$M_{0}=10$ are hereafter assumed.}

\subsection{Touch Variability}

The inter-user and inter-finger variabilities introduce an unpredictable
offset for each contacted material, notwithstanding the monotonic
behavior of the differential sensor code $\Delta s(\epsilon)$ for
a given user and gesture \cite{Bianco2020bis}. Namely, touch pressure
is expected to modify the effective permittivity sensed by the fingertip
antenna. The stronger the pressure, the larger the adhesion and hence
the higher the effective permittivity perceived by the fingertip antenna.
The precision of the measurement is hence related to the unpredictable
touch pressure.

To quantify this effect, the differential sensor code $\Delta s$
of a fingertip-sensor is measured when the touch pressure is mechanically
increased in a controlled way. For this purpose, the index finger
(diameter $1$ cm) of a volunteer is kept fixed over a bottle (diameter
$6.5$ cm) by using a plastic tie. Three pressure levels are enforced
by fastening the tie according to three perimeters: $8.5$ cm for
the low-pressure, $8$ cm for the medium, and $7.6$ cm for the high.
The corresponding pressures include the physiological variations and
range from the barely-fixed to the tightly-bound finger. The higher
the pressure applied, the better the IC adheres to the object, the
higher the returned $\Delta s$ is (Fig.~\ref{fig:Digital-fingerprints-when}).
The estimated precision is dependent on the touched object and, in
particular, it gets worse along with the increase of the permittivity.
However, the relative precision $\nicefrac{\sigma_{p}}{\Delta s}$
is bounded to $|\nicefrac{\sigma_{p}}{\Delta s}|<0.3$ so that a conservative
estimate of the precision of any measurement of $\Delta s$ with uncontrolled
touch pressure is $\sigma_{p}=$$0.3\triangle s$. However, it can
be improved (i.e., reduced) by evaluating the averaged fingerprint
$\overline{F}$ on \textit{$N_{n}$} sensors of the hand's fingertips
touching the same material as stated by the propagation of uncertainty
formula \cite{JCGM1002008,Rabinovich05}

\begin{equation}
\overline{\sigma}_{F}=\sqrt{\frac{\sum_{i=1}^{N}\sigma_{p,i}^{2}}{N_{n}^{2}}}\leq\frac{0.3}{N_{n}}\sqrt{\sum_{i=1}^{N}\Delta s_{i}^{2}}.\label{eq:sigmaFpressure}
\end{equation}
 For instance, Fig.~\ref{fig:Digital-contrasts-of-1} shows the R-FADs
digital metric and the corresponding standard deviations that are
produced by the s\textcolor{black}{ame volunteer as above without
any control on the touch pressure in the case of the only index finger,
and then when all the fingers are sensorized. By the effect of (\ref{eq:sigmaFpressure}),
the standard deviation is nearly halved in the multi-channel configuration.
It is also worth noticing that the effect of the stiffness of the
object is comparable to a different pressure applied, and that $\overline{F}$
increases by about $1$ unit every $0.6$ unit of dielectric contrast.}

\subsection{\textcolor{black}{Size and Material of the Container \label{subsec:Container-Variation}}}

\textcolor{black}{The variability of the dimension, thickness, and
material of the container of the dielectric liquids affects the sensor
code returned by the system. For instance, by varying the diameter
of the PET bottles of $\pm1.25$ cm, the uncertainty on the averaged
fingerprint is of $3$ units (Fig.~\ref{fig:Averaged-fingerprints-of}(a)).}

\textcolor{black}{A bottle of ordinary glass (diameter 6.5 cm, thickness
1.6 mm, $\epsilon=6.9$ \cite{vetro}) is considered to evaluate the
sensing capabilities of the R-FAD when using a different container.
The glass thickness and permittivity raise all the averaged fingerprints
(Fig.~\ref{fig:Averaged-fingerprints-of}(b)). Even though the contrast
between the alcohol and the deionized water is reduced, it is still
possible to distinguish the materials with a remarkable margin.}

\section{Reliability of the R-FAD Response \label{sec:Experimental-campaign}}

The last experimental campaign is aimed at quantifying the reliability
of the on-hand link when touching objects and, in particular, the
overall variability of the multi-channel R-FAD responses. A test population
of ten healthy volunteers having different hand sizes (Fig.~\ref{fig:Test-population-hand-1-1})
is sensorized over all the five fingers of the hand. They are asked
to naturally grasp (with no mechanical control on the touch pressure)
t\textcolor{black}{he three reference PET bottles as above. The} experiments
are repeated three times for a total of 450 measurements.

\subsection{Statistic of Responding Finger-Sensors}

Fig.~\ref{fig:Bar-plot-over}(a) presents the complementary cumulative
distribution (CCD) bar of the responding fingers of the same hand
so that $CDD(m)$ is the percentage, over the test population, that
at least \textit{m} fingertip sensors of the same hand respond to
the reader. Measurements show that at least one finger will provide
information in all ($CDD(1)=100\%$) combinations (user-material).
Over all the three materials, at least two fingers provide meaningful
data in 90\% of the cases and three fingers in 60\%. There is no case
when all the fingers respond at the same time.

Further insight into the communication can be gained by the polar
diagram in Fig.~\ref{fig:Bar-plot-over}(b), showing the percentage
of responding finger-sensors with respect to the filling liquid of
the bottles. The ideal condition would be all the three color markers
concentrated on the 100\% circle. No finger would always be responsive
for all the materials on the entire population of users. Finger III
is, however, the most reliable one, with a probability of 70\% of
response when touching all three objects. The thumb and the IV finger
are the least reliable ones (percentage of 20\% and 30\%, respectively).
Finally, fingers II and V provide intermediate reliability of 50\%.
Thus, sensors could be just placed on fingers II, III, and V.

\subsection{Digital Fingerprints and Statistics}

The digital fingerprints returned by the multi-channel R-FAD system
are reported in Fig.~\ref{fig:Digital-fingerprint-of} for each volunteer.
As expected by simulations, the digital fingerprints increase along
with the permittivity, and the corresponding polygons enlarge accordingly
in most of the considered cases, thus providing the base for automatic
classification of the touched materials. In some cases, up to three
finger-sensors do not respond (unfilled marke\textcolor{black}{rs),
but the presence of the other sensors permits to guarantee response
reliability at the hand level. The Kiavat diagrams show a large variability
among the sensors, that are mostly caused by }\textcolor{black}{\emph{i}}\textcolor{black}{)
the morphology of the wearer's hand and fingers, }\textcolor{black}{\emph{ii}}\textcolor{black}{)
the applied pressures, }\textcolor{black}{\emph{iii}}\textcolor{black}{)
the differences in the grasping gesture like the distance between
fingers and tags' orientation, and }\textcolor{black}{\emph{iv}}\textcolor{black}{)
the eventual fabrication imperfections of the auto-tuning ICs.}

\textcolor{black}{The inter-user va}riability of the averaged fingerprint
can be quantified with the help of the bar diagram in Fig.~\ref{fig:Averaged-fingerprints-},
wherein the responses of the populations are grouped by the kind of
materials ranked according to their permittivity. The standard deviation
is less than 11 units for all three classes and just 5 units in the
case of the lowest-permittivity liquid. These values are, however,
such to not induce overlapping among adjacent classes; therefore,
the materials can be discriminated by just applying thresholds.\textbf{ }

\section{Conclusion \textcolor{black}{and Future Works} \label{sec:Discussion-and-conclusions}}

When attached to the five fingers of the hand, the flexible and stretchable
fingertip antennas with auto-tuning RF ICs are weakly coupled, thus
enabling independent dielectric measurements. A stable RFID link with
a back-hand-mounted patch antenna interrogator can be established
with less than $25$ dBm of emitted power. A reliable measurement
of the sensor code metric, which can be related to the touched object's
permittivity, requires the calculation of the average on 10 samples
to get rid of intrinsic fluctuations and of the transient required
by auto-tuning to stabilize. The random touch force introduces a 30\%
uncertainty in the differential sensor code of the single sensor that
can be reduced to one half by resorting to a multi-sensor configuration
that also allows for a $100\%$ reliability of the communication with
at least one finger in all considered tests over a $10$-volunteer
set. The precision of the averaged fingerprint measurement is always
less than $11$ units (over the useful range $80\leq s\leq400$),
which is suitable to discriminate at least three classes of liquids
with low, medium, and high permittivity. 

\textcolor{black}{Although the method has been just applied to materials
of rather different permittivities, it is reasonably expected that
a better resolution could be achieved by the application of machine
learning classification exploiting additional information such as
the amplitude and phase of the backscattered power returned by the
device, as already done in \cite{avocado9159579}. Such algorithms
could also be exploited to recognize different grasping gestures,
allowing for sensing even when some fingersensors do not adequately
adhere or do not contact the material to be sensed. Another possible
future direction of research is the optimization of the fingertip
sensors in terms of overall size to simultaneously ensure significative
susceptance variations and similar contact areas over the five fingers
when grasping dielectric objects.}

\textcolor{black}{\bibliographystyle{IEEEtran}
\bibliography{RFAD_Characterization}
}
\end{document}